# Order-Optimal Data Aggregation in Wireless Sensor Networks – Part I: Regular Networks

Richard J. Barton, Member IEEE, and Rong Zheng, Member IEEE

*Abstract*— The predominate traffic patterns in a wireless sensor network are many-to-one and one-to-many communication. Hence, the performance of wireless sensor networks is characterized by the rate at which data can be disseminated from or aggregated to a data sink. In this paper, we consider the data aggregation problem. We demonstrate that a data aggregation rate of $\Theta(\log n / n)$ is optimal and that this rate can be achieved in wireless sensor networks using a generalization of cooperative beamforming called *cooperative time-reversal communication*.

## I. INTRODUCTION

The design and analysis of wireless sensor networks differs from that of more general data communication networks, such as the Internet or wireless mesh networks, in that the predominate traffic patterns in a sensor network are many-to-one and one-to-many communication. The performance of wireless sensor networks is thus characterized by the rate at which data can be disseminated from or aggregated to a data sink. In [3], we have investigated the broadcast capacity and information dissemination rate of multihop wireless networks. In this paper, we consider the reverse problem, called the *data aggregation problem*, which concerns the maximum sustainable rate at which each sensor can transmit data to the sink under a power constraint.

Capacity bounds for the data aggregation problem under simplified physical layer models have been established in [4-6]. In [6], the capability of large-scale sensor networks to measure and transport a two-dimensional stationary random field using sensors equipped with fixed scalar quantizers was investigated. It was shown that as the density of the sensor nodes increases to infinity, the total number of bits transmitted to the sink in order to represent the field with a given level of fidelity also increases to infinity under any compression scheme. At the same time, the single-receiver transport capacity of the network remains constant as the density increases. In [5], the more general problem of computing and communicating a symmetric function of the sensor measurements is investigated. It was shown that for a certain class of functions, called divisible functions, the maximum rate at which the function $f$ can be computed and communicated to the sink satisfies $\Theta\!\left(1 / \log\left(\left|\mathsf{R}(f,n)\right|\right)\right)$, where $n$ is the number of sensors in the network and $\mathsf{R}(f,n)$ is the range of the function $f$. Since computation of the identity function is equivalent to transporting all raw data, and the identity function is a divisible function with $\mathsf{R}(f,n) = |\mathsf{X}|^n$ for some $|\mathsf{X}| < \infty$, $\Theta(1/n)$ is a tight bound on the achievable throughput for each sensor.

Both of the studies discussed above assume a simplified protocol model for the wireless channel. In particular, each link has a fixed capacity of at most $W$, and transmissions between nodes are successful as long as other transmitters are sufficiently distant. This model clearly does not take into account the time-varying or non-deterministic nature of many channels and makes simplified assumptions regarding link capacity. For example, a Raleigh fading model is often more appropriate for links between distant nodes in an environment subject to multipath propagation, and link capacity for closely spaced network nodes is more accurately modeled as a function of signal-to-interference-and-noise ratio (SINR). Nonetheless, the $O(1/n)$ upper bound is not at all surprising, and it reflects the basic observation that for data aggregation in a multihop environment, the traffic load increases for nodes closer to the sink, and the total achievable rate is limited by the maximum rate at which the sink can receive information from its neighbors.

The authors in [4] studied the transport capacity of many-to-one dense wireless networks subject to a *total* average power constraint. It is shown that for nodes placed uniformly on a sphere of unit radius, the transport capacity of $\Theta(\log(n))$ can be achieved as the number of sensor nodes $n$ grows to infinity. This result is used to derive necessary and sufficient conditions that characterize the set of observable random fields by dense sensor networks. Our work in this paper differs from [4] in three key aspects, namely: i) nodes are individually power constrained, ii) we consider nodes dispersed regularly at a constant density in an area that

Manuscript submitted April 6, 2006. Revised and resubmitted November, 2007. R. J. Barton was supported in part by University of Houston GEAR Grant I089367. R. Zheng is supported in part by NSF CAREER award CNS-0546391. Portions of this work have been presented previously in [1, 2].
R. J. Barton is with Engineering Research and Consulting, Inc. at NASA Johnson Space Center, Houston, TX 77058 USA (phone: 281-483-1444; fax: 281-244-2327; e-mail: Richard.j.barton@nasa.gov.)
R. Zheng is with the Department of Computer Science, University of Houston, Houston TX 77204-4005 USA (e-mail: rzheng@cs.uh.edu).

increases to infinity along with the number of nodes, and iii) we explicitly take into account the effect of path losses. Therefore, the capacity of the network considered here is distance limited rather than interference limited as considered in [4].

One approach to improving the data aggregation rate in a dispersed network is to employ cooperative communication techniques, in which multiple nodes in the network cooperate to transmit information to the sink. Recently, both cooperative relay strategies [7-10] and cooperative transmission strategies [11-13] have been extensively studied. In this paper, we consider a cooperative transmission strategy based on a technique called *time-reversal communication* (TRC), which can be regarded as a generalization of beamforming. Various signal processing techniques based on time-reversal (TR) processing have been proposed and studied previously for many applications, including acoustical imaging [14], electromagnetic imaging [15], underwater acoustic communication [16], and wireless communication channels [17, 18].

In two recent publications [19, 20], we have studied some aspects of the application of TRC to cooperative communication in wireless networks. In this paper, we investigate the utility of cooperative TRC for improving the achievable data aggregation rate in wireless sensor networks. We demonstrate that a rate of $\Theta(\log n/n)$ is achievable using cooperative TRC and that this rate is in fact order-optimal for the data aggregation problem with a single-antenna sink in a fading environment.

*Main contributions*: The main contributions of this paper can best be summarized as follows:
- We establish two optimality properties for time-reversal communication and discuss their implications for the application of cooperative TRC.
- We derive a simple expression for the capacity of time-reversal communication links for fixed realizations of the Rayleigh fading channels. Using this expression, we derive an asymptotic expression for the achievable data rate of TRC links as the number of cooperating nodes goes to infinity.
- We derive the achievable asymptotic data aggregation rate using TRC and multi-hop relay and show that it is order-optimal under realistic channel models and a regular topology.
- We establish the energy efficiency of cooperative TRC and derive the asymptotic lifetime for data aggregation using TRC and multi-hop relay.

The remainder of this paper is organized as follows. In Section II, we give a brief introduction to cooperative TRC. In Section III, we present a discussion of the physical-layer and network-layer models adopted for this analysis. We derive results on achievable data aggregation rates in Section IV and corresponding results on asymptotic network lifetime in Section V. In Section VI, we present some concluding remarks. Many of the proofs have been omitted from the body of the paper and are presented in an Appendix.

## II. COOPERATIVE TIME-REVERSAL COMMUNICATION

An example of cooperative TRC is illustrated in Figure 1 below. In this case, the cluster of network nodes identified as Group *A* cooperate to transmit a common data stream to the sink node identified as the receiver in the figure. During a training phase, the sink node transmits a short sequence of wide-band training pulses which are received at all of the nodes in the cluster. After transmission of the training sequence, the receiving nodes in the cluster independently perform pulse estimation in order to estimate the exact arrival time, duration, and shape of the received pulse.

After completion of the training phase, information is transmitted from an arbitrary node in the cluster to the sink using the following cooperative TRC scheme. Whenever a node in the cluster has data to transmit to the sink node, the source node first disseminates the information to all of the other nodes in the cluster. The ensemble of nodes in the cluster then cooperate to transmit the information to the sink node by synchronously transmitting a stream of identical data symbols modulated onto the time-reverse of their respective estimated received waveforms.

The motivation for utilizing TR for communication is based primarily on the optimality properties presented in the following lemma, which are simple consequences of the Cauchy-Schwarz inequality.

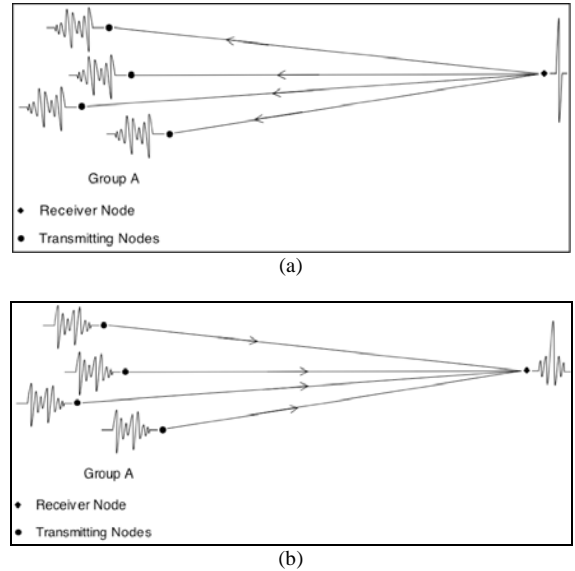

Figure 1. (a) Training for transmission from the cooperating cluster (Group *A*) to the sink (Receiver Node), (b) Transmission from the cooperating cluster to the sink.

*Lemma 1*. Let $h_i(t)$ represent the impulse response of the channel between source node *i* and the sink node, as measured at the sink node relative to time $t = 0$ at the source node, for a collection of cooperating nodes $i = 1, 2, \ldots, m$. Let $\tilde{h}_i(t) = \bar{h}_i(-t)$ represent the complex conjugate of the time reverse of the impulse response for each channel. Then, we have the following two results.

(a) The instantaneous power output from the sink node at

an arbitrary time $t = t_0$ subject to a total energy constraint at the transmitting nodes is maximized by simultaneously transmitting a constant multiple of the signal $\tilde{h}_i(t - t_0)$ from each source node at time $t = 0$. Furthermore, the received signal corresponding to the transmission of the aggregate signal $s(t) = \sum_{i=1}^{m} \tilde{h}_i(t)$ is given by the sum of the autocorrelation functions $R_{h_i}(t)$ of the individual channel impulse responses; that is,

$$r(t) = \sum_{i=1}^{m} R_{h_i}(t) = \sum_{i=1}^{m} \int_{-\infty}^{\infty} h_i(\tau) \bar{h}_i(\tau - t) d\tau.$$

(b) Let the signal transmitted from the $i^{\text{th}}$ source node be given by $s_i(t)$, and let $s(t) = \sum_{i=1}^{m} s_i(t)$ represent the aggregate transmitted signal. Let $S_i(f)$ represent the energy spectrum of the signal $s_i(t)$ and let the *space-frequency localization* of $s(t)$ be given by the quantity

$$L_s = \int_{-B/2}^{B/2} \left[ \sum_{i=1}^{m} |S_i(f)|^2 \right]^2 df,$$

where $B$ represents the two-sided bandwidth of $s(t)$. Then, the total output energy of the received signal at the sink subject to a localization constraint at the transmitting nodes is again maximized by transmitting a constant multiple of the signal $\tilde{h}_i(t - t_0)$ from each source node at time $t = 0$.

*Proof.* See Appendix.

It follows from this lemma that if perfect impulses are used in the training procedure discussed above, and if the resulting pulse estimation and synchronized time-reversed re-transmission are performed perfectly, all of the transmitted waveforms from the cooperating cluster will converge at the sink node to produce an impulsive waveform (equal to the sum of the autocorrelation functions of the various channel impulse responses) that maximizes either the peak power output from the sink at the desired time subject to an energy constraint or the total output energy at the sink subject to a localization constraint.

A few comments are warranted regarding the implications of this lemma.

- It is straightforward to show that $L_s \geq E_s^2 / B$ and that $L_s$ is minimized if $S_i(f)$ is constant for all $-B/2 \leq f \leq B/2$ and all $i = 1, 2, \ldots, m$. Conversely, for a fixed signal energy, $L_s$ can be made arbitrarily large by letting the transmitted signal be concentrated in a narrow frequency band at one transmitter. Hence, $L_s$ can be regarded as a measure of the concentration of the aggregate transmitted signal in both space and frequency. Although this type of constraint is more mathematically than physically motivated, it does imply an energy constraint, and is consistent with the physical intuition that signals should be distributed spatially at the transmitter in order to be focused spatially at the receiver (e.g., beamformer aperture vs. beamwidth) and that they should be distributed over frequency in order to be concentrated in time. That is, a constraint on the space-frequency localization at the transmitter forces a transmitted signal with fixed energy to be dispersed in both space and frequency, which is desirable for focusing the signal energy at a particular point in space and time.

- The proof of Part (a) of this lemma is identical to the proof usually given for the optimality of the matched-filter receiver for AWGN communication channels. Hence, TRC can be regarded as "matched signaling" for a channel rather than matched filtering. That is, one is matching the signal to the channel and allowing the channel itself to function as the receiver filter rather than matching the receiver filter to the channel in order to maximize the output signal-to-noise ratio at a particular sampling time.

- The lemma implies that cooperative TRC can be regarded as a generalization of cooperative or distributed beamforming that applies to both broadband and frequency-selective wireless channels where beamforming would ordinarily fail. In fact, as the bandwidth of the transmitted training pulse decreases, TRC reduces exactly to beamforming. That is, the impulse response of the each channel reduces to a single complex-valued constant.

- A side effect of optimality property (a) presented in the lemma is that the output from a TRC wireless channel tends to be concentrated in both space and time at the receiver. The extent of the temporal and spatial focusing that occurs at the receiver is determined by the spatial and temporal autocorrelation function of the channel. This is analogous to the beam pattern associated with beamforming. The analysis of the random interference process associated with cooperative TRC that leads to the physical-layer TRC channel model adopted for this study is similar in nature to the analysis of the beam pattern of randomly generated phased array beamformers studied in [13].

- In practice, of course, neither the training pulse shape, the pulse estimation, nor the transmission synchronization will be perfect, and system performance will suffer as a result. The degradation in the performance of cooperative TRC due to pulse estimation and timing errors in a non-information theoretic context has been studied previously in [19, 20]. The effect of such errors on the theoretical capacity of cooperative TRC links derived in this paper is also of interest, but remains a topic for future work.

## III. PHYSICAL-LAYER AND NETWORK-LAYER MODELS

We consider a wireless sensor network $G_n$ consisting of a group of $n$ nodes, $N = \{1, 2, \ldots, n\}$ located in the plane on a regular grid with density $\rho$. Without loss of generality, we let $\rho = 1$. Nodes are power limited, i.e., their average transmission power cannot exceed $P_{max}$. For ease of analysis, we assume nodes follow a regular layout on a grid of size $\sqrt{n} \times \sqrt{n}$.

At the physical layer, all wireless links are assumed to be baseband channels corrupted by circularly symmetric, complex-valued additive white Gaussian noise with power spectral density $N_0$ as well as additive interference from other transmitting nodes in the network, which is also assumed to be Gaussian in the aggregate. However, two different channel propagation models are assumed in our analysis depending on the nature of the communication link being analyzed.

For links in a non-cooperative multihop relay, which comprises multiple short-range communication links, we adopt a simple propagation model often used for network throughput analysis. In particular, the power on the channel is assumed to decay with distance deterministically at an exponential rate with path-loss exponent $\alpha > 2$. Hence, the maximum achievable rate (in bits/s) for communication from node $i$ to node $j$ in a non-cooperative multihop relay is given by

$$r_{ij} = B \log(1 + SINR_{ij}),$$

where $B$ represents the common bandwidth for all links on the network, $P_i$ is the power transmitted by node $i$, $\rho_{ij}$ is the distance between nodes $i$ and $j$, $I$ is the set of interfering users, and

$$SINR_{ij} = \frac{P_i \rho_{ij}^{-\alpha}}{BN_0 + \sum_{k \in I} P_k \rho_{ik}^{-\alpha}}.$$

Similarly, when common information is broadcast from node $i$ to a set of nodes $R$ in a non-cooperative multihop relay, the maximum achievable rate for the broadcast is given by

$$r_i = \min_{j \in R} \{B \log(1 + SINR_{ij})\}.$$

For analysis of cooperative TRC links, which are utilized for long-range communication, all channels are modeled as compound channels consisting of a deterministic path-loss channel with exponent $\alpha > 2$ cascaded with a Raleigh fading channel. The fading channels are assumed to be independent and identically distributed (i.i.d.) for every distinct pair of nodes $(i, j)$, and the frequency response $H_{ij}(f)$ of the channel between any two such nodes is modeled as a stationary, circularly symmetric, complex-valued Gaussian random process in the frequency domain that remains fixed during each network realization. In other words, the fading links are modeled as i.i.d. wide-sense-stationary-uncorrelated-scattering Rayleigh fading channels that are random but fixed for a very large number of packet intervals on the network. The fading processes are normalized to have unit power and coherence bandwidth $\Delta/2 > 0$, so that the autocorrelation function $\phi_{ij}(f) \equiv \phi(f)$ of the process $H_{ij}(f)$ satisfies $\phi(0) = 1$ and $\phi(f) = 0$ for all $|f| > \Delta/2$.

Adoption of two different physical layer models for short-range and long-range communication is motivated by a desire to simplify the analysis of multihop broadcast rates within a cluster of closely spaced nodes and also by the observation that communication links between nodes in close proximity are often dominated by a direct line-of-sight component and therefore exhibit less fading behavior than links between widely separated nodes. As mentioned previously, this deterministic short-range channel model is often utilized for network throughput analysis [21].

To determine the capacity for an arbitrary cooperative TRC link in our network, we assume that the cooperating cluster consists of $m \ll n$ nodes contained within a square region of side length $R$ and that the distance $d'$ from the center of the square to the sink satisfies $d'/R \gg 1$. Under this assumption, the distance $\rho_i$ from an arbitrary node $i$ in the cluster to the sink is approximately the same for all nodes in the cluster, and we can write $\rho_i \approx d'$ for all $i = 1, 2, \ldots, m$. Finally, we assume that all cooperating nodes in a cluster transmit with common average power $P \leq P_{max}$ and that while the sink is receiving a cooperative TRC transmission from an arbitrary cluster, the interference at the sink can be modeled as radiating uniformly at average power $P$ from all nodes in the network at distance greater than $\rho_0$ from the sink, where $\rho_0$ is an arbitrary constant.

## IV. ASYMPTOTIC DATA AGGREGATION RATE

We will need the following technical lemmas to prove our main result. Lemma 2 and its corollaries establish asymptotic expressions for the achievable data rate for long-range cooperative TRC links. Lemma 3 establishes an achievable broadcast rate within a cluster of closely spaced nodes using multi-hop relay. Lemma 4 establishes bounds on the amount of traffic that an arbitrary node in the network will handle in a particular multihop routing protocol.

*Lemma 2*. Under the physical-layer model discussed above for cooperative TRC with $m \gg 1$, the maximum achievable rate for an arbitrary cooperative TRC link (assuming the information has been previously broadcast to all cooperating nodes) is well modeled as

$$r_{TRC} \approx \frac{\Delta B}{\Delta + B} \log\left(1 + \frac{PX}{\frac{\Delta B}{\Delta + B} N_0 + \frac{2\pi P K_0}{(\alpha-2)\rho_0^{\alpha-2}}}\right),$$

where

$$K_0 \approx \frac{1}{B} \int_{-\Delta/2}^{\Delta/2} \mathrm{sinc}\left(\frac{2\pi(\Delta+B)f}{\Delta B}\right) \phi^2(f) df \leq \frac{\Delta}{B},$$

and $X$ is a random variable with mean $\mu_X$ and variance $\sigma_X^2$ given by

$$\mu_X \approx K_\mu (d')^{-\alpha} m^2,$$
$$\sigma_X^2 \approx K_\sigma (d')^{-2\alpha} m^3,$$

where $K_\mu$ and $K_\sigma$ are constants that are independent of $m$.

*Proof.* See Appendix.

*Corollary 1.* It follows immediately from Lemma 2 that as $m \to \infty$ we have

$$X/m^2 \xrightarrow{p} K_\mu (d')^{-\alpha},$$

where the symbol "$\xrightarrow{p}$" indicates convergence in probability. Hence,

$$r_{TRC} \approx \frac{\Delta B}{\Delta + B} \log\left(1 + \frac{m^2 K_\mu P_{\max} (d')^{-\alpha}}{\frac{\Delta B}{\Delta + B} N_0 + \frac{2\pi P K_0}{(\alpha-2)\rho_0^{\alpha-2}}}\right),$$

with high probability.

*Corollary 2.* As $B/\Delta \to \infty$, $X$ is asymptotically Gaussian with $K_\mu = 1$ and

$$K_\sigma \approx \frac{8}{B}\left[\int_0^{B/2}\left(1-\frac{f}{B}\right)\left[1-\phi^2(f)\right] {}_2F_1\left(-1,-1;1;\phi^2(f)\right) df - 1\right].$$

*Proof.* See Appendix.

*Lemma 3.* Under the physical-layer model discussed above for non-cooperative multihop relay, for any integer $k \geq 0$, there exists a time-division-multiple-access (TDMA) scheduling scheme such that one node per square of edge length $l$ can broadcast concurrently to all nodes located within a radius of $k$ squares (in Manhattan distance) with fixed rate $R(l,k)$ satisfying

$$R(l,k) \geq \frac{B}{4(k+1)^2} \log\left(1 + \frac{P_{\max}}{BN_0[l(k+1)]^\alpha + K_1 P_{\max}}\right),$$

where $K_1$ is a constant independent of $k$ and $l$.

*Proof.* See Appendix I of [3].

*Corollary 3.* It follows immediately from Lemma 3 with $l = k = 1$ that on any connected sub-graph of a grid, a total broadcast rate of $R(1,1)$ can be sustained from any number of nodes in the subgraph to all nodes in the sub-graph.

*Lemma 4.* Consider a grid with the data aggregation point $O$ at the center, as illustrated in Figure 2 below. Nodes are labeled in a 2-D coordinate system with $O$ at the origin. To route from a node $u$ located at $(x, y)$ to the aggregation point $O$, the following rules apply.

1) A Voronoi diagram is constructed on the grid. Given a set of $n$ nodes, a Voronoi tessellation is the partitioning of a plane into convex polygons such that each polygon contains exactly one generating point, and every point in a given polygon is closer to its generating point than to any other. For nodes on a grid, the Voronoi diagram is also a grid (shown in dotted lines in Figure 2).

2) Let $D_{uO}$ be a straight line from $u$ to point $O$. Let $\{u = u_0, u_1, u_2, \ldots, u_k = O\}$ be the set of nodes whose Voronoi cells intersect with $D_{uO}$.

3) The nodes $\{u_1, u_2, \ldots, u_{k-1}\}$ form the sequence of relays[2] from $u$ to $O$.

Since data is relayed through Voronoi cells with common edges, at each step, communication takes place between neighboring nodes[3] on the grid. Since each node has a unique path to $O$, data aggregation follows a tree rooted at $O$.

Let $\lambda$ be the rate of data generated by each sensor. The total amount of traffic $T(x, y)$ that a node $u$ located at $(x, y)$ must relay (including its own data) is bounded by:

$$\left\lfloor \frac{n}{\sqrt{x^2+y^2}} \cdot \frac{\sqrt{2}}{4} \right\rfloor < T(x,y) < \left\lceil \frac{n}{\sqrt{x^2+y^2}} \right\rceil, \quad (1)$$

for $1 \ll x^2 + y^2 \ll n$.

*Proof.* See Appendix.

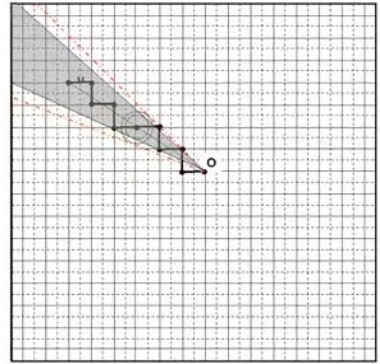

Figure 2. Routing on a Grid

Using these lemmas, we establish our main result on the achievable data aggregation rate in the following manner. To maximize the rate of data fusion at the sink, we divide the $\sqrt{n} \times \sqrt{n}$ grid into three areas as shown in Figure 3. In Areas I and III, data are aggregated using non-cooperative multihop relay on tree rooted at $O$. In Area II, nodes are organized into $R \times R$ square clusters. Data are first broadcast among all nodes inside the cluster (termed *intra-cluster communication*), then all nodes in a cluster cooperatively perform time reversal communication towards the sink (termed *inter-cluster communication*). Communication in each distinct area is carried out in *non-overlapping time slots* of equal length. This allows different communication strategies to be used without interfering with one another. The rate sustainable for each sensor in the network is determined by the minimum of the achievable rates in each area.

---

[2] In cases when node $u$ lies on the diagonals of the square area, we tilt the line $D_{uO}$ slightly to avoid intersections with the Voronoi cells at the grid points.

[3] Neighboring nodes are those that are separated by unit Manhattan distance.

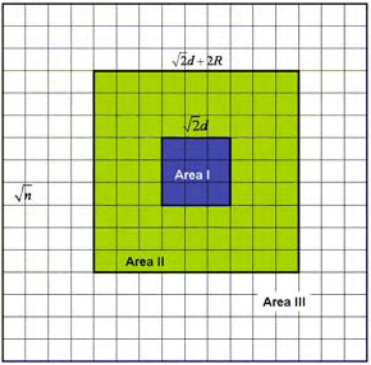

Figure 3. Partition of the network into three distinct Areas

In the following, we let $\lambda$ be the rate achievable for each sensor in the network, and we consider the collection of clusters comprising Area II with centers at distance no greater than $d' = d + R/\sqrt{2}$ from the sink. Recall that nodes are placed on regular grid with unit density. The total area of Area II is given by $\left(\sqrt{2}d + 2R\right)^2 - \left(\sqrt{2}d\right)^2 = 4\sqrt{2}dR + 4R^2$. Each cluster is a square with sides of length $R$. Thus, the number of clusters in Area II is $M = 4\left(\sqrt{2}d + R\right)/R$, and the number of nodes in each cluster is $m = R^2$. The number of nodes in Area I is $n_1 = 2d^2$, and the number of nodes in Area III is $n_3 = n - 2\left(\sqrt{2}d + 2R\right)^2$. We are interested in the rate of growth in achievable rate as $m, n \to \infty$ by including more cooperative nodes in a cluster while keeping the node density constant, and we are primarily concerned with the distance-limited rather than interference-limited regime of network operation. Hence, we use the result of Corollary 1 to characterize the asymptotic capacity of the TRC links, and we rewrite it using more compact notation as

$$r_{TRC} \approx \frac{\Delta B}{\Delta + B}\log\left(1 + \frac{m^2 d'^{-\alpha}P}{\frac{\Delta B}{\Delta + B}N_0 K'}\right) = \frac{\Delta B}{\Delta + B}\log\left(1 + \frac{R^4 d'^{-\alpha}P}{\frac{\Delta B}{\Delta + B}N_0 K'}\right),$$

where $K'$ is appropriately defined. Similarly, we rewrite the result of Lemma 2 as

$$R(l,k) = R(k) \geq \frac{B}{4(k+1)^2}\log\left(1 + \frac{P}{BN_0 K''}\right),$$

where $K''$ is appropriately defined. We can now state and prove the following theorem.

*Theorem 1*. If $2 < \alpha < 4$, the achievable data aggregation rate for the network $G_n$ as $n \to \infty$ is $\Theta(\log n/n)$.

*Proof.* The fact that the asymptotic data aggregation rate is $O(\log n/n)$ regardless of the value of $\alpha$ follows immediately from the "genie-aided" upper bound that can be derived from results in [22]. That is, in [22], it is shown that the capacity of a multiple-input, single-output fading channel with $n$ transmitting antennas, approaches $B\log\left(1 + \frac{nP}{BN_0}\right)$ for large $n$ if channel state information is available only at the receiver. A simple extension of this result gives a capacity of $B\log\left(1 + \frac{n^2 P}{BN_0}\right)$ if channel state information is also available to the transmitter. Hence, even with an infinite broadcast rate among all source nodes and all nodes at constant distance from the sink, the achievable data aggregation rate for the network would still be $O(\log n/n)$ bits/s per node.

To show that the data aggregation rate is also $\Omega(\log n/n)$ for $2 < \alpha < 4$, we began by deriving the achievable rates in each of the three areas of the network.

*Achievable Rate in Areas I and III*. In Area III, nodes forward their data towards the sink according to the above routing strategy until a node in Area II (called a *root node*) is reached. Letting $\sqrt{x^2 + y^2} = \left(\sqrt{2}d + 2R\right)/2$ in Equation (1), it follows from Lemma 4 that the traffic load for a node close to the border of Area II is upper bounded by $\left\lceil 2n/\left(\sqrt{2}d + 2R\right)\right\rceil \lambda$.

By Lemma 3 with $d' \approx d$, each root node can receive information from its closest neighbor at rate $R(1) = \frac{B}{16}\log\left(1 + \frac{P}{BN_0 K''}\right)$. Hence, since communication in each area is carried out in independent time slots, the sustainable rate in Area III satisfies

$$\left\lceil \frac{2n}{\sqrt{2}d + 2R}\right\rceil \lambda \geq \frac{B}{48}\log\left(1 + \frac{P}{BN_0 K''}\right). \qquad (2)$$

Similarly, in Area I, by the same argument, we have

$$(2d - R)^2 \lambda \geq \frac{B}{48}\log\left(1 + \frac{P}{BN_0 K''}\right). \qquad (3)$$

*Achievable Rate in Area II*. From Corollary 1, we know that each cluster of size $R^2$ can transmit at rate $\frac{\Delta B}{\Delta + B}\log\left(1 + \frac{R^4 (d')^{-\alpha}P}{\frac{\Delta B}{\Delta + B}N_0 K'}\right)$ to a node at distance $d'$. Using TDMA to separate transmissions in each of the clusters as well as each of the three areas, the *effective rate* of inter-cluster communication for each cluster is thus

$$r_{inter} = \frac{\Delta B}{3M(\Delta + B)} \log\left(1 + \frac{R^4 (d')^{-\alpha} P}{\frac{\Delta B}{\Delta + B} N_0 K'}\right)^4.$$ On the other hand, assuming that we separate intra-cluster and inter-cluster communication into different time slots and allow all clusters to perform intra-cluster communication concurrently, the intra-cluster broadcast rate is given by $r_{intra} = \frac{B}{48} \log\left(1 + \frac{P}{BN_0 K''}\right)$, as in Area I. If $M \to \infty$, it follows that the total effective achievable rate for each cluster is given by

$$\frac{r_{intra} r_{inter}}{r_{intra} + r_{inter}} \approx r_{inter} = \frac{\Delta B}{3M(\Delta + B)} \log\left(1 + \frac{R^4 (d')^{-\alpha} P}{\frac{\Delta B}{\Delta + B} N_0 K'}\right).$$

Now, since the nodes in Area II must transport all of the traffic from Area III to the sink, the amount of traffic that each cluster must carry is no greater than

$$\lambda R^2 + \left\lceil \frac{2n}{\sqrt{2}d + 2R} \right\rceil 2R\lambda \approx \lambda R^2 + \frac{16n\lambda}{M+4},$$

where the first term gives the amount of traffic generated from the cluster itself, and the second term corresponds to the aggregated load from Area III. If we let $d = n^\beta$, $R = n^\gamma$, with $0 < \gamma < \beta < \frac{1}{2}$, then as $n \to \infty$, we have $M \approx 4\sqrt{2} n^{\beta-\gamma}$. It follows that the sustainable rate in Area II satisfies

$$\lambda R^2 + \frac{16n\lambda}{M+4} \approx \frac{16n}{M} \lambda \geq \frac{\Delta B}{3M(\Delta + B)} \log\left(1 + \frac{R^4 (d')^{-\alpha} P}{\frac{\Delta B}{\Delta + B} N_0 K'}\right),$$

or equivalently

$$\lambda \geq \frac{\Delta B}{48n(\Delta + B)} \log\left(1 + \frac{R^4 (d')^{-\alpha} P}{\frac{\Delta B}{\Delta + B} N_0 K'}\right), \quad (4)$$

as $n \to \infty$.

*Achievable Rate for the Network.* Comparing Equations (2)-(4) and noting that $d' \approx d$ as $n \to \infty$, we see that the entire network can sustain a rate of

$$\Omega\left(\frac{\Delta B}{48n(\Delta + B)} \log\left(1 + \frac{R^4 (d')^{-\alpha} P}{\frac{\Delta B}{\Delta + B} N_0 K'}\right)\right).$$

Furthermore, if $\alpha < 4$ and we choose $0 < \gamma < \beta < \frac{4}{\alpha}\gamma$ with $\beta < \frac{1}{2}$, then we can achieve a sustainable rate of $\Omega(\log n / n)$ bit/s per node, as claimed. ∎

## V. Asymptotic Network Lifetime for Data Aggregation

Time reversal is of interest not only because it can be utilized to achieve higher data aggregation rates in an extended wireless network, but also because it improves the energy efficiency of the network. In this section, we study asymptotic network lifetime, defined as the time at which power is exhausted at the first node. For analytical purposes, we assume that the network operates in a low SNR regime with low duty cycle. In particular, we assume the data generation rate at each sensor is $\lambda = o(1/n)$. We further assume that the sink node is equipped with an external power source and would therefore not exhaust its power supply.

*Non-cooperative multihop data fusion.* We first compute the power consumption of a baseline naïve non-cooperitive multihop data fusion scheme, in which data are aggregated on a tree rooted at the data aggregation point. Using the Shannon formula, the power consumption required to transmit at rate $\gamma$ to a node at unit distance in the noise-limited regime is given by $BN_0 \left(e^{\gamma/B} - 1\right)$. Nodes closest to the sink carry the highest traffic load, which takes the form $c_2 n\lambda$ for some constant $0 < c_2 < 1$. Hence, it follows from the assumption $\lambda = o(1/n)$ that, as $n \to \infty$, the maximum power consumption is given by

$$BN_0 \left(e^{c_2 n\lambda/B} - 1\right) \approx c_2 n\lambda N_0.$$

Consequently, the asymptotic network lifetime is given by $\frac{e_0}{c_2 n\lambda N_0}$ second, where $e_0$ is the initial energy.

*Cooperative transmission using TRC.* Sensors in Area II must forward traffic from Area III in addition to their local traffic. Therefore, the power consumption for sensors in Area II is necessarily higher than for sensors in Area III.

Now, consider an arbitrary sensor in Area II. It must perform two operations, as discussed in Section IV above. First, during intra-cluster communication, it broadcasts (or forwards) a load of $\lambda R^2 + \frac{16n\lambda}{M+4}$ to all other nodes in the same cluster. Since non-cooperative multihop relay is used at this stage, the corresponding power consumption is given by

$$\frac{\Delta B}{\Delta + B} N_0 \left[ e^{\frac{\Delta + B}{\Delta B}\left(\lambda R^2 + \frac{16n\lambda}{M+4}\right)} - 1 \right] \approx N_0 \left(\lambda R^2 + \frac{16n\lambda}{M+4}\right).$$

Second, using Corollary 1 with $m = R^2$ (see Equation (4)), the power consumption during inter-cluster communication using cooperative TRC, is given by

$$\frac{\Delta B K' N_0 \left(e^{48n\lambda(\Delta+B)/\Delta B} - 1\right)}{(\Delta + B) R^4 (d')^{-\alpha}} \approx \frac{48n\lambda K' N_0}{R^4 (d')^{-\alpha}}.$$

If $\alpha < 4$, we choose $0 < \gamma < \beta < \frac{4}{\alpha}\gamma$ with $\beta < \frac{1}{2}$, $d = n^\beta$, $R = n^\gamma$, and $M = 4\sqrt{2} n^{\beta-\gamma}$. Hence, as $n \to \infty$, the total power consumption at a node in Area II is given by

$$P_{II} = \frac{48n\lambda K'N_0}{R^4 d'^{-\alpha}} + \frac{16n\lambda N_0}{M}$$
$$= \lambda N_0 \left( \lambda R^2 + \frac{16n}{M} + \frac{48K'n}{R^4 d'^{-\alpha}} \right) \quad (5)$$
$$= \lambda N_0 n \left( n^{2\gamma-1} + 16n^{-(\beta-\gamma)} + 48K'n^{-(4\gamma-\alpha\beta)} \right).$$

---

[4] In fact, since we are using TDMA to separate the cluster transmissions, there will be no interference at the sink, and the constant $K'$ in this expression actually satisfies $K' = 1$. Since this does not affect the result, we continue to use the more general expression.

In Area I, sensors close to the sink have the highest traffic load, which is given by $(2d-R)^2 \lambda$ as $n \to \infty$. Therefore, the power consumption in Area I is given by

$$P_I = BN_0 \left[ e^{(2d-R)^2 \lambda / B} - 1 \right] \approx N_0 n^{2\beta} \lambda. \quad (6)$$

Putting all of this together, it follows from Equations (5) and (6) that the network lifetime is given by

$$\frac{e_0}{\max(P_I, P_{II})} = \frac{e_0}{c_3 n \lambda} \cdot n^{\min(1-2\beta, 4\gamma-\alpha\beta, \beta-\gamma)}, \quad (7)$$

where $c_3$ is a properly defined constant. Clearly, under the conditions $\alpha < 4$, $0 < \gamma < \beta < \frac{4}{\alpha}\gamma$, and $\beta < \frac{1}{2}$, the exponent $\min(1-2\beta, 4\gamma-\alpha\beta, \beta-\gamma)$ is a positive number. Hence, compared with non-cooperative multihop relay, cooperative TRC can improve the network lifetime significantly and thus is indeed more energy efficient.

## VI. CONCLUSION

We have established that an asymptotic data aggregation rate of $\Theta(\log n/n)$ is optimal for a wireless sensor network operating in a fading environment with a power path-loss exponent that satisfies $2 < \alpha < 4$. Furthermore, we have shown that this rate can be achieved using a new cooperative transmission scheme called time-reversal communication in conjunction with a novel data aggregation protocol.

In addition, we have demonstrated that the asymptotic lifetime of a sensor network operating in the same fading environment in the low SNR, low duty-cycle regime can be increased significantly using cooperative TRC together with the data aggregation protocol proposed herein for large extended networks.

The optimal asymptotic data aggregation rate for the high-attenuation case, when, $\alpha > 4$, has not been considered here. This case is treated in detail in Part II of this paper, which also generalizes the results presented here to randomly placed rather than regularly placed network nodes. It is shown in Part II that the optimal data aggregation rate when $\alpha > 4$ is $\Theta(\log n/n)$ and that this rate can be achieved using a non-cooperative multihop relay strategy.

## APPENDIX

*Proof of Lemma 1.* To prove Part (a), let $s(t) = \sum_{i=1}^{m} s_i(t)$ represent an arbitrary aggregate transmitted signal that satisfies $\mathsf{E}_s = \sum_{i=1}^{m} \int_{-\infty}^{\infty} |s_i(t)|^2 dt \leq \mathsf{E}_{\max}$ for some energy constraint $\mathsf{E}_{\max}$. Then, the received signal at time $t = t_0$ satisfies

$$|r(t_0)|^2 = \left| \sum_{i=1}^{m} \int_{-\infty}^{\infty} s_i(t) h_i(t_0 - t) dt \right|^2$$
$$\leq \left[ \sum_{i=1}^{m} \int_{-\infty}^{\infty} |s_i(t)|^2 dt \right] \left[ \sum_{i=1}^{m} \int_{-\infty}^{\infty} |h_i(t_0 - t)|^2 dt \right]$$
$$= \mathsf{E}_s \left[ \sum_{i=1}^{m} \int_{-\infty}^{\infty} |h_i(t)|^2 dt \right]$$
$$\leq \mathsf{E}_{\max} \left[ \sum_{i=1}^{m} R_{h_i}(0) \right],$$

with equality if and only if $\mathsf{E}_s = \mathsf{E}_{\max}$ and $s_i(t) = c\overline{h}_i(t_0 - t) = c\tilde{h}_i(t - t_0)$ for all $(i,t)$ and some constant $c > 0$. Further, if $s(t) = \sum_{i=1}^{m} \tilde{h}_i(t)$, then it is clear that $r(t)$ takes the stated form. This proves Part (a).

To prove Part (b), let $H_i(f)$, $i = 1, 2, \ldots, m$, represent the frequency response for each of the node-sink channels, and let the aggregate transmitted signal satisfy

$$\mathsf{L}_s = \int_{-B/2}^{B/2} \left[ \sum_{i=1}^{m} |S_i(f)|^2 \right] df \leq \mathsf{L}_{\max},$$

for some space-frequency localization constraint $\mathsf{L}_{\max}$. Then, the total output energy of the received signal at the sink is given by

$$\mathsf{E}_r = \int_{-B/2}^{B/2} \left| \sum_{i=1}^{m} S_i(f) H_i(f) \right|^2 df,$$

and

$$E_r = \int_{-B/2}^{B/2} \left| \sum_{i=1}^{m} S_i(f) H_i(f) \right|^2 df$$

$$= \int_{-B/2}^{B/2} \sum_{i=1}^{m} \sum_{j=1}^{m} S_i(f) \bar{S}_j(f) H_i(f) \bar{H}_j(f) df$$

$$\leq \sqrt{\int_{-B/2}^{B/2} \left[ \sum_{i=1}^{m} |H_i(f)|^2 \right]^2 df} \sqrt{\int_{-B/2}^{B/2} \left[ \sum_{i=1}^{m} |S_i(f)|^2 \right]^2 df}$$

$$= \sqrt{L_s \int_{-B/2}^{B/2} \left[ \sum_{i=1}^{m} |H_i(f)|^2 \right]^2 df}$$

$$\leq \sqrt{L_{max} \int_{-B/2}^{B/2} \left[ \sum_{i=1}^{m} |H_i(f)|^2 \right]^2 df},$$

with equality if and only if $L_s = L_{max}$ and $S_i(f)\bar{S}_j(f) = c\bar{H}_i(f)H_j(f)$ for all $(i,j,f)$ and some $c > 0$. In particular, equality is achieved by choosing $s_i(t) = \sqrt{c}\tilde{h}_i(t-t_0)$, as before. This proves Part (b) and establishes the lemma. ∎

*Proof of Lemma 2.* Let the average power $P_{max}$ transmitted by all nodes satisfy $P_{max} = \Delta B E/(\Delta + B)$, where $E$ represents the average transmitted energy per symbol for each node, and $\Delta B/(\Delta + B)$ represents an estimate of the maximum symbol rate achievable on a fading link without incurring intersymbol interference.[5] Let $m$ represent the number of nodes in a cluster and let $H_i(f)$, $i = 1, 2, \ldots, m$, represent the (random) frequency response of the WSSUS fading channel connecting node $i$ to the sink. Finally, let $r_i \approx r$ represent the (approximately constant) distance from each node in the cluster to the sink. Then the energy spectrum of the TRC signal transmitted at time $t = 0$ by node $i$ is given by

$$S_i(f) = \sqrt{\frac{E}{E_i B}} e^{-i2\pi f t_0} \bar{H}_i(f),$$

where

$$E_i = \frac{1}{B} \int_{-B/2}^{B/2} |H_i(f)|^2 df,$$

and the energy spectrum of the aggregate signal received at the sink corresponding to the cooperative TRC transmission of a single symbol from the cluster at time $t = 0$ is given by

$$S(f) = \sqrt{\frac{E}{B}} e^{-i2\pi f t_0} \sum_{i=1}^{m} \frac{r_i^{-\alpha/2}}{\sqrt{E_i}} |H_i(f)|^2$$

$$\approx \sqrt{\frac{E}{B}} e^{-i2\pi f t_0} r^{-\alpha/2} \sum_{i=1}^{m} \frac{1}{\sqrt{E_i}} |H_i(f)|^2.$$

Assuming that a matched filter receiver is implemented at the sink, the output from the receiver at time $t = t_0$, excluding additive noise, is given by

$$S = \sqrt{\frac{E}{B}} \int_{-B/2}^{B/2} \sum_{i,j=1}^{m} r_i^{-\alpha/2} r_j^{-\alpha/2} |H_j(f)|^2 |H_i(f)|^2 df$$

$$\approx \sqrt{\frac{E}{B}} r^{-\alpha} \int_{-B/2}^{B/2} \sum_{i,j=1}^{m} \frac{1}{\sqrt{E_i E_j}} |H_j(f)|^2 |H_i(f)|^2 df$$

$$= \sqrt{P_{max}(1 + B/\Delta)X},$$

where

$$X = \frac{r^{-\alpha}}{B} \int_{-B/2}^{B/2} \sum_{i,j=1}^{m} \frac{1}{\sqrt{E_i E_j}} |H_j(f)|^2 |H_i(f)|^2 df.$$

To compute the interference at the sink, we assume that an arbitrary node in the network located at distance $\rho$ from the sink transmits a random signal with energy spectrum

$$e^{-i2\pi f \tau} \bar{G}_1(f)$$

where $\bar{G}_1(f)$ has the same distribution as the WSSUS fading channels on the network and $\tau$ is uniformly distributed over the interval $[-(\Delta+B)/\Delta B, (\Delta+B)/\Delta B]$. If that signal propagates to the sink over a fading channel with frequency response $G_2(f)$, then the additive interference at the output from the TRC matched filter at the sink at time $t = t_0$ is given by

$$I(\rho) \approx \sqrt{\frac{E}{B}} (r\rho)^{-\alpha/2} \left[ \int_{-B/2}^{B/2} e^{i2\pi f(t_0 - \tau)} \bar{G}_1(f) G_2(f) \cdot \left( \sum_{i=1}^{m} \frac{1}{\sqrt{E_i}} |H_i(f)|^2 \right) df \right].$$

Using the results on distributions of functions of Gaussian random variables tabulated in [23], it is straightforward to show that the mean value of $I(\rho)$ is zero and the variance is given by

$$\sigma_I^2(\rho) \approx (r\rho)^{-\alpha} \frac{E}{B} \left[ \begin{array}{c} \int_{-B/2}^{B/2} \int_{-B/2}^{B/2} \text{sinc}\left(\frac{2\pi(\Delta+B)(f-\lambda)}{\Delta B}\right) \\ \cdot \phi^2(f - \lambda) \\ \cdot \sum_{i,j=1}^{m} \frac{1}{\sqrt{E_i E_j}} |H_j(f)|^2 |H_i(\lambda)|^2 df d\lambda \end{array} \right].$$

Summing the contribution of all such independent random interference components uniformly distributed around the sink at distances $\rho \geq \rho_0$ and assuming that $H_i(f) \approx H_i(\lambda)$ for $|f - \lambda| \leq \Delta/2$, we see that as the network grows infinitely large, the total interference at the sink can be modeled as a Gaussian random variable $I$ with mean zero and variance

$$\sigma_I^2 = \int_{\rho_0}^{\infty} \sigma_I^2(\rho) d\rho$$

$$\approx \frac{2\pi P_{max}(1 + B/\Delta)}{(\alpha - 2)\rho_0^{\alpha-2}}$$

$$\cdot \left[ \frac{r^{-\alpha}}{B} \int_{-B/2}^{B/2} \sum_{i,j=1}^{m} \frac{1}{\sqrt{E_i E_j}} |H_j(f)|^2 |H_i(f)|^2 df \right.$$

$$\left. \cdot \frac{1}{B} \int_{-\Delta/2}^{\Delta/2} \text{sinc}\left(\frac{2\pi(\Delta + B)\nu}{\Delta B}\right) \phi^2(\nu) d\nu \right]$$

$$= \frac{2\pi P_{max}(1 + B/\Delta) K_0}{(\alpha - 2)\rho_0^{\alpha-2}} X,$$

where $X$ is as defined above and

$$K_0 = \frac{1}{B} \int_{-\Delta/2}^{\Delta/2} \text{sinc}\left(\frac{2\pi(\Delta + B)\nu}{\Delta B}\right) \phi^2(\nu) d\nu.$$

---

[5] The symbol duration $T_s$ (without the delay spread) satisfies $T_s \geq 1/B$. The delay spread satisfies $\tau_d \approx 1/\Delta$. Hence, the total symbol interval $T$ satisfies $T \geq 1/B + 1/\Delta = (\Delta + B)/(\Delta B)$ and the corresponding symbol rate $R$ satisfies $R = 1/T \leq (\Delta B)/(\Delta + B)$.

Proceeding in a similar fashion, it is straightforward to show that the additive background noise at the output of the receiver is a Gaussian random variable $N$ with mean zero and variance $\sigma_N^2 = BN_0 X$, where $N_0$ is the power spectral density of the AWGN process at the input to the receiver.

Finally, for $m \gg 1$, we have

$$\mu_X \approx \frac{r^{-\alpha}}{B} \sum_{i \neq j=1}^{m} \int_{-B/2}^{B/2} E\left\{ \frac{1}{\sqrt{E_i E_j}} |H_i(f)|^2 |H_j(f)|^2 \right\} df$$

$$= \frac{r^{-\alpha}}{B} \sum_{i \neq j=1}^{m} \int_{-B/2}^{B/2} E\left\{ \frac{1}{\sqrt{E_i}} |H_i(f)|^2 \right\} E\left\{ \frac{1}{\sqrt{E_j}} |H_j(f)|^2 \right\} df$$

$$\approx \frac{r^{-\alpha}}{B} m^2 \int_{-B/2}^{B/2} \left( E\left\{ \frac{1}{\sqrt{E_1}} |H_1(f)|^2 \right\} \right)^2 df,$$

and

$$\sigma_X^2 \approx \frac{r^{-2\alpha}}{B^2} \sum_{i \neq j \neq k=1}^{m} \int_{-B/2}^{B/2} \int_{-B/2}^{B/2} E\left\{ \frac{1}{\sqrt{E_i E_j E_i E_k}} \cdot |H_i(f)|^2 |H_j(f)|^2 \right.$$
$$\left. \cdot |H_i(\lambda)|^2 |H_k(\lambda)|^2 \right\} df d\lambda$$

$$- \frac{r^{-2\alpha}}{B^2} \sum_{i \neq j \neq k=1}^{m} \int_{-B/2}^{B/2} E\left\{ \frac{1}{\sqrt{E_i E_j}} |H_i(f)|^2 |H_j(f)|^2 \right\} df$$

$$\cdot \int_{-B/2}^{B/2} E\left\{ \frac{1}{\sqrt{E_i E_k}} |H_i(\lambda)|^2 |H_k(\lambda)|^2 \right\} d\lambda$$

$$\approx \frac{r^{-2\alpha}}{B^2} m^3 \int_{-B/2}^{B/2} \int_{-B/2}^{B/2} E\left\{ \frac{1}{E_1} |H_1(f)|^2 |H_1(\lambda)|^2 \right\}$$

$$\cdot E\left\{ \frac{1}{\sqrt{E_1}} |H_1(f)|^2 \right\} E\left\{ \frac{1}{\sqrt{E_1}} |H_1(\lambda)|^2 \right\} df d\lambda$$

$$- \frac{r^{-2\alpha}}{B^2} m^3 \left[ \int_{-B/2}^{B/2} \left( E\left\{ \frac{1}{\sqrt{E_1}} |H_1(f)|^2 \right\} \right)^2 df \right]^2.$$

Hence,

$$\mu_X \approx K_\mu m^2 r^{-\alpha},$$
$$\sigma_X^2 \approx K_\sigma m^3 r^{-2\alpha},$$

where

$$K_\mu = E\left\{ \frac{1}{\sqrt{E_1}} |H_1(f)|^2 \right\},$$

and

$$K_\sigma = \frac{1}{B} \int_{-B/2}^{B/2} \int_{-B/2}^{B/2} E\left\{ \frac{1}{E_1} |H_1(f)|^2 |H_1(\lambda)|^2 \right\} df d\lambda - K_\mu^2.$$

It follows that the output of the receiver at the sink takes the form

$$Y = \sqrt{P_{max}(1 + B/\Delta)} X + N + I,$$

where $N \sim \mathbb{N}(0, \sigma_N^2)$ and $I \sim \mathbb{N}(0, \sigma_I^2)$, so the (random) capacity of the TRC link is given by

$$r_{TRC} = \frac{\Delta B}{\Delta + B} \log\left( 1 + \frac{P_{max}(1 + B/\Delta) X^2}{BN_0 X + \frac{2\pi P(1 + B/\Delta) K_0}{(\alpha - 2)\rho_0^{\alpha-2}} X} \right)$$

$$= \frac{\Delta B}{\Delta + B} \log\left( 1 + \frac{P_{max} X}{\frac{\Delta B}{\Delta + B} N_0 + \frac{2\pi P K_0}{(\alpha - 2)\rho_0^{\alpha-2}}} \right),$$

where $X$ is a random variable with mean $\mu_X$ and variance

$\sigma_X^2$ given above. ∎

*Proof of Corollary 2.* Using the results on distributions of functions of Gaussian random variables tabulated in [23], it is straightforward to show that the mean and variance of the random quantity $\mathsf{E}_i$ for $i = 1, 2, \ldots, m$ are given by

$$\mu_{\mathsf{E}_i} = 1,$$

$$\sigma_{\mathsf{E}_i}^2 = \frac{2\Delta}{B} \int_0^{B/\Delta} \left(1 - \frac{\Delta}{B}u\right)\left[1 - \phi^2(\Delta u)\right]^4 {}_2F_1\left(2,2;1;\phi^2(\Delta u)\right) du - 1$$

$$= \frac{2\Delta}{B} \int_0^{1/2} \left(1 - \frac{\Delta}{B}u\right)\left[1 - \phi^2(\Delta u)\right]^4 {}_2F_1\left(2,2;1;\phi^2(\Delta u)\right) du - \frac{\Delta}{B} + \left(\frac{\Delta}{2B}\right)^2,$$

where ${}_2F_1(a,b;c;z)$ represents the hypergeometric function (see, for example, [24]). Under the assumption that $B/\Delta \to \infty$, we have $\sigma_{\mathsf{E}_i}^2 \to 0$ uniformly for all $i = 1, 2, \ldots, m$. Hence, for sufficiently large values of $B/\Delta$, we have $\mathsf{E}_i \approx 1$ for all $i = 1, 2, \ldots, m$, and the random variable $X$ is well approximated as

$$X = \frac{r^{-\alpha}}{B} \int_{-B/2}^{B/2} \sum_{i,j=1}^{m} |H_j(f)|^2 |H_i(f)|^2 df.$$

For the remainder of this analysis, we assume that $X$ takes this slightly simplified form. It then again follows from the results in [23] that $K_\mu$ and $K_\sigma$ take the simplified forms $K_\mu = 1$ and

$$K_\sigma = \frac{8}{B}\left[\int_0^{B/2}\left(1 - \frac{f}{B}\right)\left[1 - \phi^2(f)\right] {}_2F_1\left(-1,-1;1;\phi^2(f)\right) df - 1\right],$$

as claimed.

Finally, to see that $X$ is approximately Gaussian as $B/\Delta \to \infty$, notice that the random variable $X$ can be rewritten as

$$X = \frac{r^{-\alpha}}{B} \int_{-B/2}^{B/2} \sum_{i,j=1}^{m} |H_j(f)|^2 |H_i(f)|^2 df$$

$$= \frac{r^{-\alpha}}{B} \left[\sum_{k=0}^{2B/\Delta - 1} \int_{-B/2+k\Delta/2}^{-B/2+(k+1)\Delta/2} \sum_{i,j=1}^{m} |H_j(f)|^2 |H_i(f)|^2 df\right].$$

That is, $X$ is the sum of a sequence of $2B/\Delta$ two-dependent identically distributed random variables. Hence, invoking an $M$-dependent version of the central limit theorem (see, for example, [25]), as $B/\Delta \to \infty$, the random variable $X$ is well approximated as a Gaussian random variable. ∎

*Proof of Lemma 4.* Consider two concentric disks centered at $u$ with radius $\sqrt{2}/2$ and 1 respectively. Clearly, the Voronoi cell containing $u$ is fully contained in the outer disk and completely contains the inner disk. Therefore:

1) The sufficient condition for a node $v$ to route through $u$ is that $v$ is further from $O$ than $u$ and falls in the cone formed by $O$, the two tangent lines of the inner disk, and the boundary of the $\sqrt{n} \times \sqrt{n}$ grid (indicated by the shaded area in figure 2), which we denote by $A$.

2) The necessary condition for a node $v$ to route through $u$ is that $v$ is is further from O than $u$ and falls in the cone formed by $O$, the two tangent lines of the outer disk, and the boundary of the $\sqrt{n} \times \sqrt{n}$ grid, which we denote by $B$.

Hence, to determine the traffic load at node $u$, it is equivalent to compute the number of nodes falling into the regions $A$ and $B$ defined above. For ease of computation, we simplify the boundary of $\sqrt{n} \times \sqrt{n}$ grid using the circumscribed and inscribed circles. We have,

$$\text{Area}(A) > \left(\frac{n}{x^2+y^2} - 1\right)\sqrt{x^2+y^2-\frac{1}{2}} \cdot \frac{\sqrt{2}}{4},$$

and

$$\text{Area}(B) < \left(\frac{2n}{x^2+y^2} - 1\right)\sqrt{x^2+y^2-1} \cdot \frac{1}{2}.$$

When $1 \ll x^2 + y^2 \ll n$, the traffic load $T(x,y)$ at $u$ satisfies:

$$\left\lfloor \frac{n}{\sqrt{x^2+y^2}} \cdot \frac{\sqrt{2}}{4} \right\rfloor \lambda < T(x,y) < \left\lceil \frac{n}{\sqrt{x^2+y^2}} \right\rceil \lambda.$$

This result demonstrates that for nodes at constant distance to the sink, the traffic loads differ *at most* by a constant factor (i.e., $\sqrt{2}/4$) under the proposed routing strategy. In other words, the proposed routing is load balanced. ∎

REFERENCES

[1] R. J. Barton and R. Zheng, "Order-Optimal Data Aggregation in Wireless Sensor Networks Using Cooperative Time-Reversal Communication," in *Proceedings of 40th Annual Conference on Information Sciences and Systems (CISS)*, Princeton, NJ, 2006, pp. 1050-1055.

[2] R. J. Barton and R. Zheng, "The Impact of Time-Reversal Modulation on the Performance of Cooperative Relaying Strategies in Wireless Networks," in *Proceedings of Information Theory and Applications Workshop*, San Diego, CA, 2006.

[3] R. Zheng, "Information Dissemination in Power-Constrained Wireless Networks," in *Proceedings of 25th Annual Joint Conference of the IEEE Computer and Communications Societies (INFOCOM)*, 2006.

[4] H. El Gamal, "On the Scaling Laws of Dense Wireless Sensor Networks: The Data Gathering Channel," *IEEE Transactions on Information Theory,* vol. 51, pp. 1229-1234, March 2005.

[5] A. Giridhar and P. R. Kumar, "Computing and Communicating Functions over Sensor Networks," *IEEE Journal on Selected Areas in Communications,* vol. 23, pp. 755-764, April 2005.

[6] D. Marco, E. J. Duarte-Melo, M. Liu, and D. L. Neuhoff, "On the Many-to-One Transport Capacity of a Dense Wireless Sensor Network and the Compressibility of Its Data.," in *Proceedings of Second International Workshop on Information Processing in Sensor Networks (IPSN)*, 2003.

[7] P. Gupta and P. R. Kumar, "The Capacity of Wireless Networks," *IEEE Transactions on Information Theory,* vol. 46, pp. 388-404, March 2000.

[8] G. Kramer, M. Gastpar, and P. Gupta, "Cooperative Strategies and Capacity Theorems for Relay Networks," *IEEE Transactions on Information Theory,* vol. 51, pp. 3037-3063, Sept. 2005.

[9] R. U. Nabar, H. Bolcskei, and F. W. Kneubuhler, "Fading Relay Channels: Performance Limits and Space-Time Signal Design," *IEEE Journal on Selected Areas in Communications,* vol. 22, pp. 1099-1109, August 2004.

[10] A. Sendonaris, E. Erkip, and B. Aazhang, "User Cooperation Diversity - Part Ii: Implementation Aspects and Performance Analysis," *IEEE Transactions on Communications,* vol. 51, pp. 1939-1948, Nov. 2003.

[11] G. Barriac, R. Mudumbai, and U. Madhow, "Distributed Beamforming for Information Transfer in Sensor Networks," in *Proceedings of Third International Symposium on Information Processing in Sensor Networks*, 2004, pp. 81-88.

[12] P. Mitran, H. Ochiai, and V. Tarokh, "Space-Time Diversity Enhancements Using Collaborative Communications," *IEEE Transactions on Information Theory,* vol. 51, pp. 2041-2057, June 2005.

[13] H. Ochiai, P. Mitran, H. V. Poor, and V. Tarokh, "Collaborative Beamforming for Distributed Wireless Ad Hoc Sensor Networks," *IEEE Transactions on Signal Processing,* vol. 53, pp. 4110-4124, Nov. 2005.

[14] P. Roux, A. Derode, A. Peyre, A. Tourin, and M. Fink, "Acoustical Imaging through a Multiple Scattering Medium Using a Time-Reversal Mirror," *Journal of the Acoustical Society of America,* vol. 107, pp. L7-L12, Feb. 2000.

[15] A. B. Ruffin, J. Van Rudd, J. Decker, L. Sanchez-Palencia, L. Le Hors, J. F. Whitaker, and T. B. Norris, "Time Reversal Terahertz Imaging," *IEEE Journal of Quantum Electronics,* vol. 38, pp. 1110-1119, August 2002.

[16] G. F. EdelMann, W. S. Hodgkiss, S. Kim, W. A. Kupeman, and H. C. Song, "Underwater Acoustic Communication Using Time Reversal," in *Proceedings of MST/IEEE OCEANS Conference and Exhibition*, 2001, pp. 2231-2235.

[17] P. Kyritsi, G. Papanicolau, P. Eggers, and A. Oprea, "MISO Time Reversal and Delay-Spread Compression for FWA Channels at 5 GHz," *IEEE Antennas and Wireless Propagation Letters,* vol. 3, pp. 96-99, 2004.

[18] T. Strohmer, M. Emami, J. Hansen, G. Papanicolau, and A. J. Paulraj, "Application of Time-Reversal with MMSE Equalizer to UWB Communications," in *Proceedings of Globecom 2004*, 2004, pp. 3123-3127.

[19] R. J. Barton, J. Chen, and K. Huang, "Cooperative Time Reversal for Communication in Power-Constrained Wireless Sensor Networks," in *Proceedings of Forty-Third Annual Allerton Conference on Communication, Control, and Computing*, Monticello, IL, 2005.

[20] R. J. Barton, J. Chen, K. Huang, and D. Wu, "Cooperative Time-Reversal Communication in Wireless Sensor Networks," in *Proceedings of IEEE Workshop on Statistical Signal processing*, Bordeaux, France, 2005.

[21] M. Yuksel and E. Erkip, "Diversity-Multiplexing Tradeoff in Cooperative Wierless Systems," in *Proceedings of 40th Conference on Information Sciences and Systems*, Princeton, NJ, 2006, pp. 22-24.

[22] I. E. Telatar, "Capacity of Multi-Antenna Gaussian Channels," AT&T Bell Labs, Technical Memo 1995.

[23] M. K. Simon, *Probability Distributions Involving Gaussian Random Variables - a Handbook for Engineers and Scientists*. Boston: Kluwer Academic Publishers, 2002.

[24] M. Abramowitz and I. A. Stegun, "Handbook of Mathematical Functions," New York: Dover Publishers, Inc., 1965.

[25] R. J. Serfling, "Contributions to Central Limit Theory for Dependent Variables," *Annals of Mathematical Statistics,* vol. 39, pp. 1158-1175, Aug. 1968.